\documentclass[twocolumn,showpacs,preprintnumbers,superscriptaddress,amsmath,amssymb]{revtex4}
\usepackage{epsfig}
\usepackage{subfigure}
\usepackage{amsmath}
\usepackage{amssymb}
\usepackage{graphicx}% Include figure files
\usepackage{dcolumn}% Align table columns on decimal point
\usepackage{bm}% bold math
\input epsf

\begin{document}

\title{Parameter mismatch estimation using large deviations from synchronization}

\author{Jupiter Bagaipo}
\email{jbagaipo@mail.umd.edu}
\affiliation{
Department of Physics,
University of Maryland, College Park, Maryland 20742, USA
}
\affiliation{
Department of Mathematics,
University of Maryland, College Park, Maryland 20742, USA
}

\author{Juan G. Restrepo}
\affiliation{
Department of Mathematics,
University of Maryland, College Park, Maryland 20742, USA
}
\affiliation{
Institute for Research in Electronics and Applied Physics, 
University of Maryland, College
Park, Maryland 20742, USA
}

\date{\today}

\begin{abstract}
We present a method to determine the relative parameter mismatch in a collection of nearly identical
chaotic oscillators by measuring large deviations from the synchronized state.
We demonstrate our method with an ensemble of slightly different circle maps. 
We discuss how to apply our method when there is noise, and show an example where the noise intensity is comparable to the mismatch.
\end{abstract}


\maketitle

The study of networks of coupled dynamical systems is an important area of research 
with applications in diverse fields, ranging from biology to laser physics \cite{newman1}-\cite{pikovsky}. 
The synchronization of coupled oscillators has been under extensive study in recent years and, 
in particular, the synchronization of 
identical oscillators  has received considerable interest \cite{pikovsky,mosekilde}. Since it is impossible 
in practice to obtain identical oscillators, the effect of the difference in the parameters of the oscillators,
or {\it parameter mismatch}, might be relevant in some applications. It might be desired to 
have dynamical units as similar to each other as possible,
or to know the characteristics of the parameter mismatch in a collection of nearly identical systems.
In this paper we propose a method to use deviations from synchronization to extract information on
the parameter mismatch of the coupled dynamical units. Existing methods for parameter estimation 
(see, for example, Refs.~\cite{parlitz}-\cite{tao}) usually rely on knowledge of the typically small synchronization error. 
Our method depends on relatively large deviations from the synchronized state, and might be useful in cases in which 
the small synchronization error can not be measured accurately.

When a number of identical systems are appropriately coupled in a network, a solution exists in 
which the state of all oscillators at all times is the same. This is referred to as 
{\it identical} synchronization. This concept is useful only when the systems are identical. We will
deal with systems that are nearly, but not exactly, identical. 
We will refer to a situation in which the states of the systems are very close to each other
as {\it nearly identical} synchronization.
A method to determine the stability of the synchronous state when the systems are identical, the {\it master stability function}, 
has been proposed by Pecora and Carroll \cite{pecora2}.

In the case of nearly identical chaotic systems, the nearly synchronized state might be interrupted 
by relatively short periods of desynchronization ({\it desynchronization bursts}). 
These bursts develop with spatial patterns on the network. These spatial patterns, and the parameters for which 
they can be expected, can be predicted from the Laplacian matrix describing the network connections, 
the master stability function of the attractor, and the unstable periodic orbits embedded in it \cite{ours}. 
The spatial patterns of the bursts depend on the parameter mismatch
of the different systems. We use this fact to infer the relative deviation of the parameters of 
the individual units with respect to their mean from the desynchronization
bursts. The proposed method is as follows. The oscillators are connected in such a way that the parameter 
mismatch determines the spatial patterns of the desynchronization bursts. As we will see later, 
one such way is all-to-all coupling. The system is set up in a parameter region in which desynchronization 
bursts are expected. While a burst is developing, measurements are taken of the deviations of the 
different systems from the synchronous state. From these observations, 
the relative deviations of the parameters from the mean are deduced.

Some limitations of this method are the following. It is assumed that measurements can be taken with enough
precision such that the deviations from the synchronous state can be measured in the linear regime. 
Also, the presence of noise affects the spatial patterns of the bursts. Although we will describe 
how to deal with the noise, the effectiveness of the method decreases as the ratio of noise to mismatch increases. 
It is also assumed that unavoidable small differences in the way in which the systems 
are connected to each other does not introduce a difference 
between the systems which is of the same order of magnitude or larger than the parameter mismatch being measured.

In Sec.~I we briefly describe the master stability function method and its extension to deal with 
nearly identical systems.
In Sec.~II we present and illustrate our method with an example in the case where the noise is negligible.
In Sec.~III we discuss how to deal with the noise and show an example. In Sec.~IV we present our conclusions.

\section{Background}

For simplicity, we will use one dimensional maps for the chaotic units. The results 
generalize to other dynamical systems \cite{pecora2,ours}.
We consider a model system of $N$ dynamical units, each
one of which, when isolated, satisfies $X^{i}_{n+1}  = F(X^{i}_{n},\mu_i)$, 
where $X_{n}^{i}$ is the value of unit $i$ at time $n$ and $\mu_i$ is a 
parameter vector for system $i$.

The systems, when coupled, are taken to satisfy (e.g., \cite{pecora2}) 
\begin{equation}\label{eq:coupled}
X^{i}_{n+1}  = F(X^{i}_{n},\mu_i) - g Z\left(\sum_{j = 1}^{N}G_{ij}H(X^j)\right),
\end{equation}
where $Z$ is a function such that $Z(0) = 0$,
$G$ is a symmetric Laplacian matrix ($\sum_{j} G_{ij} = 0$) describing the network
connections, and $H$ is a function independent of $i$ and $j$. [In our examples, 
we will take $Z(x) = \sin(2\pi x)$.]
The constant $g$ determines the strength of the coupling.

If the systems are identical (i.e., $\mu_i = \mu$ for all $i$), 
there is an exactly synchronized
solution of Eqs.~(\ref{eq:coupled}), $X_n^{1} \equiv X_n^{2} \equiv \dots \equiv X_n^{N} = s_n$, 
whose time evolution is the same
as the uncoupled dynamics of a single unit, $s_{n+1} = F(s_n)$, where $F(s) = F(s,\mu)$.
The stability of the synchronized state can be determined from
the variational equations obtained by considering an infinitesimal
perturbation $\delta^{i}$ from the synchronous state, $X^{i}_{n}
= s_n + \delta^{i}_n$,
\begin{equation}\label{eq:linearized}
\delta_{n+1}^{i} = DF(s_n)\delta_{n}^{i} - g Z'(0)\sum_{j = 1}^{N}
G_{ij} DH(s_n)\delta_{n}^{j}.
\end{equation}
Let $\delta = [\delta^{1},\delta^{2},\dots,\delta^{N}]$,
and define  the vector $\eta =
[\eta^{1},\eta^{2},\dots,\eta^{N}]$ by $\delta = \eta L^{T}$, where
$L$ is the orthogonal matrix whose
columns are the corresponding real orthonormal eigenvectors of $G$; $G L = L \Lambda$,
$\Lambda = diag(\lambda_{1},\lambda_{2},\dots,\lambda_{N})$,
 where $\lambda_{k}$ is the eigenvalue of $G$ for
eigenvector $k$. Then Eqs.~(\ref{eq:linearized}) are equivalent to
\begin{equation}\label{eq:componentwise}
\eta_{n+1}^{k} = \left[DF(s_n) - g Z'(0)\lambda_{k} DH(s_n)\right]\eta_{n}^{k}.
\end{equation}
The quantity $\eta^{k}$ is the weight of the $k$th eigenvector
of $G$ in the perturbation $\delta$.  The linear stability of
each `spatial' mode $k$ is determined by the stability of the
solution of Eq.~(\ref{eq:componentwise}).
By introducing a scalar variable $\alpha = g Z'(0) \lambda_{k}$, the set
of equations given by Eq.~(\ref{eq:componentwise}) can be encapsulated
in the single equation,
\begin{equation}\label{eq:master}
\eta_{n+1} = \left[DF(s_n) - \alpha DH(s_n)\right]\eta_{n}.
\end{equation}
The master stability  function $\Psi (\alpha)$ \cite{pecora2} is the
largest Lyapunov exponent for this equation. This function depends only on the
coupling function $H$ and the dynamics of an individual
uncoupled element, but not on the network connectivity. The
network connectivity determines the eigenvalues $\lambda_{k}$
(independent of details of the dynamics of the chaotic units). 
The stability of the
synchronized  state of the network is determined by $\Psi_{*} =
\sup_{k} \Psi(g \lambda_{k})$, where $\Psi_{*}>0$ indicates
instability. 

If the systems are slightly different, one gets instead of Eqs.~(\ref{eq:linearized}) the equations
\begin{equation}
\delta^{i}_{n+1}  = D\overline{F}(s_{n})\delta_n^{i} - 
g Z'(0)\sum_{j = 1}^{N} G_{ij}DH(s) \delta_{n}^{j} + Q^{i}(s_n),
\end{equation}
where $\overline{F}(X^i) \equiv  F(X^{i}, \sum_i^N \mu_i/N)$, 
and $Q^{i}(X^{i}) \equiv F(X^{i},\mu_i) - \overline{F}(X^{i})$ represents the effect
of the mismatch and is assumed to be small. Terms of order $Q\delta$ were neglected. 
Defining $Q = [Q^{1}(s_n),Q^{2}(s_n),\dots,Q^{N}(s_n)]$, 
we obtain an equation analogous to Eq.~(\ref{eq:componentwise}),
\begin{equation}\label{eq:misma}
\eta_{n+1}^{k} = \left[DF(s_n) - g Z'(0)\lambda_{k} DH(s_n)\right]\eta_{n}^{k} + (Q L)^{k},
\end{equation}
where $(Q L)^{k}$ is the kth element of the vector $Q L$.
The Lyapunov exponent for the solution of Eq.~(\ref{eq:componentwise}) is $h_k = \Psi(g\lambda_k)$.
Assuming a solution of Eq.~(\ref{eq:misma}) to have the same average damping as that 
for Eq.~(\ref{eq:componentwise}), then, if $h_k$ is negative for all modes, 
the amplitude of $\eta_k$ can be estimated as
\begin{equation}\label{eq:machete}
\langle\left|\eta^{k}\right|\rangle 
\sim \frac{\langle \left|(QL)^{k}\right|\rangle}{1-e^{-\left|h_{k}\right|}}.
\end{equation}
(For example, if we model Eq.~(\ref{eq:misma}) by the simple system $\eta_{n+1}=e^{-h}\eta_n + q$, 
then $\eta_n$ satisfies, as $n \rightarrow \infty$, $\eta \rightarrow \frac{q}{1-e^{-h}}$. See \cite{ours}.)
The largest Lyapunov exponent $h_k$ above corresponds to a typical trajectory in the chaotic attractor. 
However, the Lyapunov exponent
for unstable periodic orbits embedded in the attractor might be larger. 
Assume that one of these periodic orbits has 
a positive Lyapunov exponent and the attractor has a negative Lyapunov exponent.
In this case, most of the time the amplitude of $\eta_k$ will be very small and given 
approximately by Eq.~(\ref{eq:machete}).
Eventually the trajectory $s_n$ will get very close to this transversally unstable periodic orbit. While 
it is close to this orbit, $\eta$ in 
Eq.~(\ref{eq:misma}) is no longer damped and gets exponentially amplified 
with the Lyapunov exponent
of the unstable periodic orbit. The deviation from the synchronized state becomes 
large, producing a
desynchronization burst. If there are no other attractors, the system returns 
to the synchronized state 
and the process repeats. Desynchronization bursts can be expected when the master 
stability function for a typical
trajectory is negative for all modes, and there is an embedded unstable periodic 
orbit which has a positive master 
stability function for at least one mode (i.e., $\overline{\Psi}(g\lambda_k) > 0$ for some $k$, where
$\overline{\Psi}$ is the master stability function of one of the embedded periodic orbits).

We will now use the fact that modes with the same eigenvalue have the same stability. 
For simplicity, assume that the coupling is all to all, so that all the modes (except the 
mode in the synchronization manifold, which has zero eigenvalue) have the same eigenvalue, $\lambda_k = N$. 
Eq.~(\ref{eq:machete}) implies that the coefficients in the eigenvector decomposition of the deviations from 
synchrony ($\eta^{k}$) are, on average, proportional to those for the deviations of the mismatch 
parameters from their mean [$(QL)^{k}$]. 
It follows that the mismatch vector $Q$ is proportional to the vector $\delta_n$ while these
approximations are valid. 
If the vector $\delta_n$ is measured, the deviations of the mismatch from its mean 
can be determined approximately up to an unknown scaling factor.

For this method to work, the measurements need to be made when the system is still in the linear regime. 
Since it is assumed that there is a limitation in the measurement accuracy, $\delta_n$ needs to be small
enough to guarantee linear behavior, but large enough to be measured. One can thus set up the system
so that desynchronization bursts are expected and make measurements while a 
desynchronization burst is developing. If the
system allows continuous tuning of the coupling strength, one could also increase it so that the 
synchronous state becomes unstable and take measurements as the system desynchronizes.

\section{Parameter mismatch without noise}

\begin{figure}[ht]
\begin{center}
\epsfig{file = 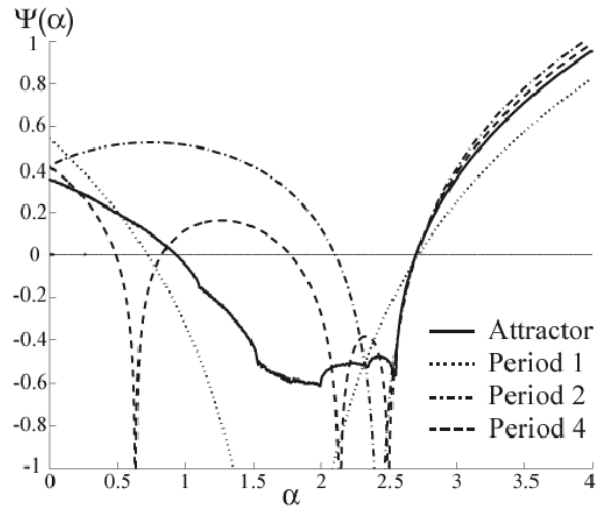, clip =  ,width=0.9\linewidth }
\caption{Master stability function, $\Psi(\alpha)$, for a typical trajectory in the attractor (continuous curve), 
for the period 1 orbit (dotted curve), for the period 2 orbit (dashed-dotted curve), and for the period 4 orbit (dashed curve).}
\label{fig:msf}
\end{center}
\end{figure}

To illustrate our method, we use the \emph{circle map}, described by the equation
\begin{equation}\label{eq:map}
\theta_{n+1}=[\theta_n + \omega + \kappa \sin 2 \pi \theta_n]\mbox{ mod } 1.
\end{equation}
We choose the parameters to be $\omega=\frac{\sqrt{5}-1}{2}$ and $\kappa=\frac{1}{\sqrt{3}}$. 
These parameters produce a chaotic attractor in $\theta \in [0.21,0.47]$. 
We found the embedded periodic orbits up to period four. To determine the orbits of period $p$ we used Newton's method 
to find the roots of $\theta=f^p(\theta)$, where $f(\theta)$ is described by Eq.~(\ref{eq:map}) and $f^p$ denotes the $p$ 
times composition of $f$. Eliminating all the orbits outside of the attractor, we found one period 1 orbit, 
two period 2 orbits, and one period 4 orbit. We show in Fig.~\ref{fig:msf} the master stability functions 
of the orbits found, and the master stability function of the attractor. Here $\alpha = 2 \pi g \lambda_k$, 
where $\lambda_k$ is the $k$th eigenvalue of the coupling matrix and $g$ is the global strength of the coupling. 

For definiteness, we assume a network that is coupled \emph{all to all}. This means that for a network of $N$ systems, 
an element in the coupling matrix $G_{ij}$ is given by
\begin{equation}\label{eq:matrix}
G_{ij} = \left\{ \begin{array}{cl}
         N-1 & \mbox{if $i=j$;} \\
          -1  & \mbox{if $i \neq j$.}
         \end{array} \right.
\end{equation}
This matrix has two distinct eigenvalues, $\lambda_0=0$ and $\lambda_k= N$ for $k=1,2,\ldots,N-1$. We ignore the 
0 eigenvalue since this corresponds to a perturbation in which all of the systems are displaced by the same amount 
(thus they remain synchronized). Due to the 
lack of other distinct eigenvalues, it is easy to pick an $\alpha$ that produces desynchrozination bursts. 
For our map the region where bubbling is expected is where $0.9 < 2\pi g \lambda_k < 2.1$ 
(note that for our example $\lambda_k$ is the same for all $k$). 

We now present an example of the method where noise is negligible. 
The mismatch is chosen to be in $\kappa$ since it has 
a more complicated effect than mismatch in $\omega$. 
The coupled systems can then be described by the general equation
\begin{equation}\label{eq:nonoise}
\theta_{n+1}^i=[\theta_n^i + \omega+(\kappa+\delta\kappa^i)\sin 2 \pi \theta_n^i - \Phi^i_n] \mbox{ mod } 1,
\end{equation}
where $\delta\kappa^i$ is the mismatch in system $i$, 
$\Phi^i_n = g \sin\left( 2 \pi \sum_{j=1}^N G_{ij} \theta_n^j\right)$, and $i,j=1,2,\ldots,N$ are indices 
representing the $i$th and $j$th system in the network [cf. Eq.~(\ref{eq:coupled})]. The term $\Phi$ 
determines the coupling of the network. We chose $N=5$ systems, $g=0.0477$ as the global coupling, 
and $\delta\kappa=[4,-1,2,-6,-2] \times 10^{-6}$ as the values for our examples. 
It should, however, be noted that this method works for any number of systems (with $g$ being adjusted accordingly) 
and mismatch of any size, although, if the mismatch becomes small, the waiting time for a desynchronization burst becomes large. 
The waiting time can be adjusted by changing the values of $N$, $g$, and $\delta\kappa$. 
\begin{figure}[ht]
\begin{center}
\epsfig{file = 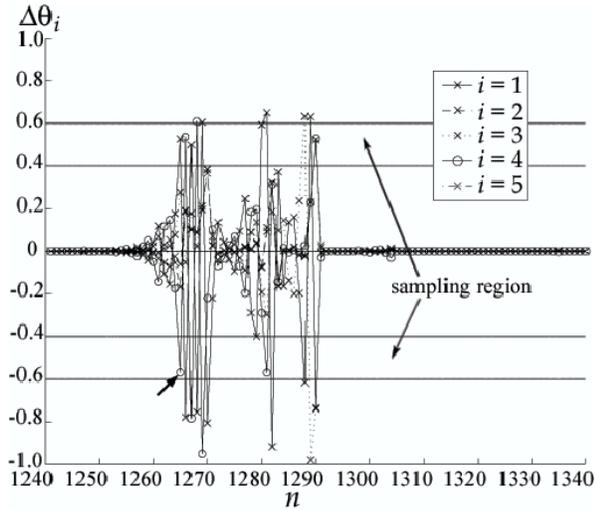, clip =  ,width=0.9\linewidth }
\caption{Plot of $\Delta \theta_n^i$ vs. $n$ near a region of desynchronization burst. 
The arrow points to the maximum of $\Delta \theta_n^i$ that is within the sampling region.}
\label{fig:bub1}
\end{center}
\end{figure}

Defining $\Delta \theta_n^i=\sin 2 \pi (\theta_n^i - \bar{\theta}_n)$, 
where $\bar{\theta}_n=\frac{1}{N}\sum_{i=1}^N \theta_n^i$, we plot $\Delta \theta_n^i$ versus $n$ 
and look for desynchronization bursts in the network.  In Fig.~\ref{fig:bub1} we show the time 
evolution of $\Delta \theta_n^i$ near a desynchronization burst. 
Our interest is in the vector $\theta_l$ where $l$ is the first time that $max_i\{\theta^i_n\}$ 
is in the \emph{sampling region}, defined as $0.4<|\Delta \theta_n^i|<0.6$ (see Fig.~\ref{fig:bub1}). 
This region is determined by the limitations on the ability to accurately measure $\Delta\theta_n^i$ 
and the dynamics of the system considered. 
The latter exists because the master stability function method relies on the systems being close to synchronization. 
During the desynchronization burst, the difference in the systems can be so large that the linearization used 
in the analysis of Sec.~I no longer applies. 
We determined that the upper bound to this region in the circle map is $|\Delta \theta_n^i|\approx 0.6$ 
or $|\theta_n^i - \bar{\theta}_n| \approx 0.10$. 
The lower bound was arbitrarily chosen as representing the accuracy of the measurements, 
which we assume is not enough to measure the mismatch directly. 
Generally the method becomes more effective the smaller the lower bound is.

\begin{figure}[ht]
\begin{center}
\epsfig{file = 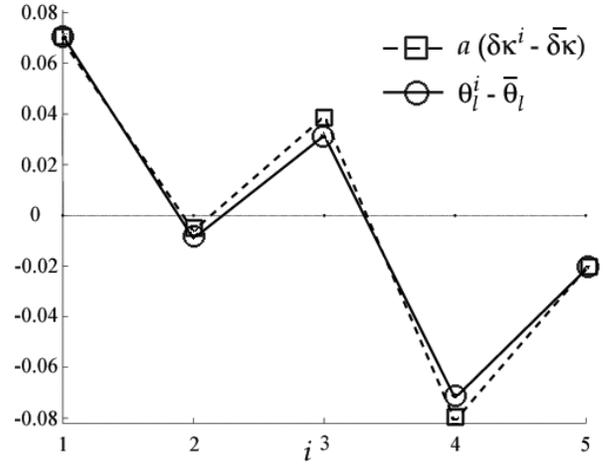, clip =  ,width=0.9\linewidth }
\caption{Superimposed plot of $a(\delta\kappa^i-\bar{\delta\kappa})$ (dashed line with square markers) and $\theta_l^i-\bar{\theta}_l$ 
(solid line with circle markers) versus $i$ with $a \approx 1.5\times10^4$.}
\label{fig:secII}
\end{center}
\end{figure}

According to the previous section, at time $l$ we should have approximately $\theta_l^i-\bar{\theta}_l \propto \delta\kappa^i-\bar{\delta\kappa}$, 
where $\bar{\delta\kappa}=\frac{1}{N}\sum_{i=1}^N \delta\kappa^i$. 
We can then obtain the relative deviations of the mismatch parameters, $\delta\kappa^i$, 
by measuring the much larger values of $\theta_l^i-\bar{\theta}_l$.
In Fig.~\ref{fig:secII} we show a superimposed plot of $\theta^i_l-\bar{\theta}_l$ 
and $a(\delta\kappa^i-\bar{\delta\kappa})$ versus $i$ where $a$, the scaling factor, minimizes
 $\sum_{i=1}^N [(\theta_l^i - \bar{\theta}_l)-a(\delta \kappa^i - \bar{\delta \kappa})]^2$. In Fig.~\ref{fig:secII} 
we calculated $a\approx 1.5\times10^4$ and this corresponds 
to the amplification of the mismatch. It should be noted that the sign of $a$ is undetermined unless we 
have knowledge of $\delta\kappa^i$. We see from the figure that 
$a(\delta\kappa^i - \bar{\delta\kappa}) \approx \theta_l^i-\bar{\theta}_l$.

The definition of the sampling region is somewhat arbitrary, and it may occur that nonlinear effects still play a role
in the resulting spatial pattern of the burst. In fact, in Fig.~\ref{fig:secII} we observe that there are still small deviations 
from the real mismatch pattern. In order to take this into account, we can take the average over various bursts. 
In the next section, we will discuss how to appropriately take the average, and we will also deal with the effects of noise.

\section{Parameter mismatch with noise}

After learning from the simpler model in Sec.~II, we can now analyze a more realistic situation. 
The method proposed and explained in the previous section applies to a similar network with noise, 
but there are a few adjustments to be made. We use the same model described by Eq.~(\ref{eq:nonoise}), except we modify it to
\begin{equation}\label{eq:noisy}
\theta_{n+1}^i=[\theta_n^i + \omega+(\kappa+\delta\kappa^i)\sin 2 \pi \theta_n^i
-\Phi_n^i+ \epsilon_n^i]\mbox{ mod } 1,
\end{equation}
where $\delta\kappa^i$, $\Phi_n^i$ are defined in the same way as before, and $\epsilon^i_n$ is a random 
variable uncorrelated at different $i$ and $n$ and simulates the noise. 
In our example, we choose $\epsilon^i_n$ uniformly from the interval $[-10^{-5},10^{-5}]$ (note that 
the noise and mismatch $\delta \kappa$ are of comparable size).

As mentioned in the previous section, nonlinear effects might produce deviations from the simple 
relation $\theta_l^i-\bar{\theta}_l=a(\delta\kappa^i-\bar{\delta\kappa})$. 
We assume that the effects of the nonlinearity and the noise can be represented by a random variable $\sigma^i$, 
such that $\theta_l^i-\bar{\theta}_l=a(\delta\kappa^i-\bar{\delta\kappa})+\sigma^i$. 
We furthermore assume that $\sigma^i$ has zero mean. Under these assumptions, the mismatch is given by
\begin{equation}\label{eq:etadiff}
\delta\kappa^i-\bar{\delta\kappa}=\left\langle \frac{\theta_l^i-\bar{\theta}_l}{a} \right\rangle - 
\left\langle \frac{\sigma^i}{a} \right\rangle,
\end{equation}
where the brackets represent an average over realizations of $\sigma$.
Because of the definition of the sampling region, the scaling factors for different samples will have similar magnitude, 
but possibly different sign. We thus get approximately, assuming the sign of $a$ is independent of $\sigma^i$,
\begin{equation}\label{eq:etascale}
\delta\kappa^i-\bar{\delta\kappa} \propto \langle \mbox{sign}(a)(\theta_l^i-\bar{\theta}_l) \rangle.
\end{equation}
Since the sign of $a$ is unknown, we use a \emph{least-squares optimization} to find the signs which minimize 
the dispersion from the mean. More precisely, if we have $M$ samples of $\theta_l$'s, we can define an average 
to be $\bar{\Theta}=\frac{1}{M} \sum_{m=1}^M \beta_m \theta_m$, where $\{\beta_m\}$ is a sequence of $1$'s and $-1$'s 
and $\theta_m$ is the vector $[\theta_l^i-\bar{\theta}_l]$ for the $m$th sample. Note that $\bar{\Theta}$ is an $N$-dimensional 
vector and that its $i$th component is an average of the $i$th component of the $M$ samples of $\theta_l$'s. 
If we then minimize the error, defined by
\begin{equation}\label{eq:err}
\mbox{error}=\frac{1}{M}\sum_{m=1}^M \|\beta_m \theta_m - \bar{\Theta}\|^2,
\end{equation}
we can find an optimal sequence of $1$'s and $-1$'s, which we shall call $\beta^*$, that ensures most of 
the $\theta_l$'s are oriented the same way.

To minimize this error and optimize $\beta$ it is impractical to test all possibilities since there are $2^M$ different $\beta$'s. 
We can, however, follow an algorithm starting with a randomly generated $\beta$. At each iterate, we generate three new $\beta$'s. 
The first, $\beta_1$, is a new random sequence, $\beta_2$ is $\beta$ altered such that the signs of 1\% of the sequence are changed, 
and $\beta_3$ 
is defined in a similar way but with 5\% of the signs changed. We can then compare $\beta_1$, $\beta_2$, $\beta_3$, and $\beta$ and 
determine which one has the smaller error determined by Eq.~(\ref{eq:err}). The one with the smallest error is then redefined as $\beta$ 
and the process is repeated until an approximation to $\beta^*$, which we denote as $\bar{\beta}$, is found. 
We can then define $\Theta_*=\frac{1}{M}\sum_{m=1}^M \bar{\beta}_m \theta_m$.
\begin{figure}[ht]
\begin{center}
\epsfig{file = 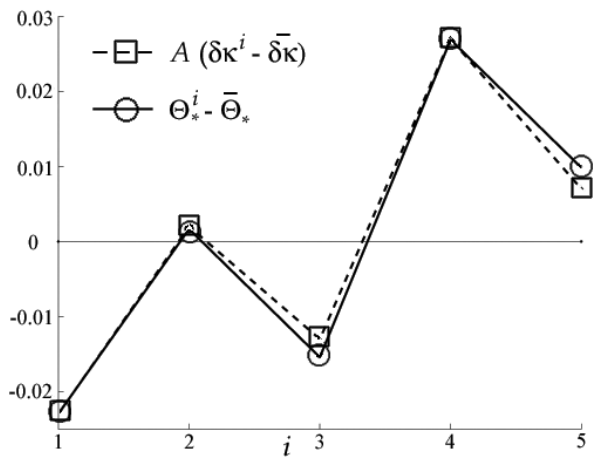, clip =  ,width=0.9\linewidth }
\caption{Superimposed plot of $A(\delta\kappa^i-\bar{\delta\kappa})$ (dashed line with square markers) and $\Theta_*^i-\bar{\Theta}_*$ 
(solid line with circle markers) versus $i$ with $A \approx -5\times10^3$.}
\label{fig:secIII}
\end{center}
\end{figure}

We show in Fig.~\ref{fig:secIII} a superimposed plot of $\Theta^i_*-\bar{\Theta}_*$ and $A(\delta\kappa^i-\bar{\delta\kappa})$ versus $i$ where $A$ minimizes $\sum_{i=1}^N [(\Theta_*^i-\bar{\Theta}_*) - A(\delta \kappa^i-\bar{\delta \kappa})]^2$. 
To obtain $\Theta_*$ we repeated the process of optimization for $M=1000$ $10^6$ times. According to the 
discussion above, we should have approximately $\Theta_*^i-\bar{\Theta}_* \propto \delta\kappa^i -\bar{\delta\kappa}$ 
where $\bar{\Theta}_*=\frac{1}{N}\sum_{i=1}^N \Theta^i_*$. 
Indeed we see that $A(\delta\kappa^i-\bar{\delta\kappa}) \approx \Theta_*^i-\bar{\Theta}_*$ even when the noise 
was comparable to the mismatch [$\epsilon \approx \delta \kappa$ in Eq.~(\ref{eq:noisy})].

\section{Discussion}

We have presented a method to use large deviations from synchronization in order to determine the 
characteristics of the parameter mismatch in a collection of nearly identical chaotic dynamical systems.
It has been noted that knowledge and manipulation of the mismatch patterns can be advantageous in order to improve
the quality of the synchronization \cite{ours}. 
The main advantage of our method is that it only requires direct knowledge of the synchronization
error when it is large enough to be measured. Furthermore, in principle, there are no limitations 
on the number of systems it can handle. 
On the other hand, the method only provides the relative deviations from the mean
and has yet to be extended to systems with comparable mismatch in different parameters.  

We have demonstrated our method by determining the relative parameter mismatch in an ensemble of $5$ circle maps.
By measuring the large deviations from the synchronized state that occur during a desynchronization burst, we were
able to determine the very small relative differences in parameters (see Fig.~\ref{fig:secII}).
We considered the presence of noise, and dealt with it by suitably averaging the measurements taken
for various desynchronization bursts. For a noise comparable to the mismatch we were able to
determine the relative parameter mismatch by averaging $1000$ realizations (see Fig.~\ref{fig:secIII}). 
For both situations, we were able to determine the relative parameter mismatch from 
measurable values even when the mismatch itself was assumed to be immeasurable.

Acknowledgements: We thank Professor Ed Ott for encouraging and useful discussions.
This work was sponsored by ONR (Physics) and by NSF (contracts PHYS 0098632 and DMS 0104087).


\begin{thebibliography}{99}

\bibitem{newman1} M.E.J. Newman, SIAM Review {\bf 45}, 167 (2003). 
%Review on networks

\bibitem{barabasi1}
A.-L. Barab\'{a}si, and R. Albert, Rev. Mod. Phys. {\bf 74}, 47 (2002).

\bibitem{pikovsky} A. Pikovsky, M.G. Rosenblum, and J. Kurths,
{\it Synchronization: A universal concept in nonlinear sciences},
(Cambridge University Press, Cambridge, 2001).
%'Fundamentals of synchronizationin chaotic systems, concepts, and applications'

\bibitem{mosekilde} E. Mosekilde, Y. Maistrenko, and D. Postnov, 
{\it Chaotic Synchronization: Applications to Living Systems} (World Scientific,
Singapore, 2002).
%book on the subject

\bibitem{parlitz} U. Parlitz, Phys. Rev. Lett. {\bf 76}, 1232 (1996).

\bibitem{maybhate} A. Maybhate and R. E. Amritkar, Phys. Rev. E {\bf 61}, 6461 (2000).

\bibitem{sakaguchi} H. Sakaguchi, Phys. Rev. E {\bf 65}, 027201 (2002).

\bibitem{tao} C. Tao, Y. Zhang, G. Du, and J.J. Jiang, Phys. Rev. E {\bf 69}, 036204 (2004).


\bibitem{pecora2} L.M. Pecora and T.L. Carroll, Phys. Rev. Lett. {\bf 80}, 2109 (1998).
%msf

\bibitem{ours} J.G. Restrepo, E. Ott, and B.R. Hunt, {\bf 69}, 066215 (2004).  



\end{thebibliography}
\end{document}